# Internalizing Null Extraterrestrial "Signals":
# An Astrobiological App for a Technological Society

Eric J. Chaisson
Harvard-Smithsonian Center for Astrophysics
Harvard University, Cambridge, Massachusetts USA

One of the beneficial outcomes of searching for life in the Universe is that it grants greater awareness of our own problems here on Earth. Lack of contact with alien beings to date might actually comprise a null "signal" pointing humankind toward a viable future. Astrobiology has surprising practical applications to human society; within the larger cosmological context of cosmic evolution, astrobiology clarifies the energetic essence of complex systems throughout the Universe, including technological intelligence that is intimately dependent on energy and likely will be for as long as it endures. The "message" contained in the "signal" with which today's society needs to cope is reasonably this: Only solar energy can power our civilization going forward without soiling the environment with increased heat yet robustly driving the economy with increased per capita energy usage. The null "signals" from extraterrestrials also offer a rational solution to the Fermi paradox as a principle of cosmic selection likely limits galactic civilizations in time as well as in space: Those advanced life-forms anywhere in the Universe that wisely adopt, and quickly too, the energy of their parent star probably survive, and those that don't, don't.

## The Context

A few years ago, I had the pleasure of attending the 50[th] anniversary of Project Ozma—the first dedicated search for extraterrestrial intelligence (SETI) conducted by Frank Drake in 1960. The celebratory gathering was held at the National Radio Astronomy Observatory in Green Bank, West Virginia, where that initial search was attempted and where I had in the intervening years operated dozens of radio-frequency experiments of my own, including a few unauthorized reality-checks for signs of otherworldly life. Although I never detected there any signal implying contact, I often wondered why not. Astronomers are commissioned by society to keep our eyes on the sky, yet we have never found any unambiguous, confirmed evidence for life beyond Earth. Are alien civilizations out there but not advanced enough to betray their presence? Or are they so advanced they are actively hiding from us? Perhaps they just don't exist at all, thereby ensuring that we are alone in the observable Universe.

I have always imagined the SETI search parameter space to resemble a vast chandelier containing some billion light bulbs, each representing a star with a habitable planet orbiting it in the Milky Way Galaxy. I cannot recall who first conceived this analogy; it probably wasn't me although I've used it for decades in my classes and writings to frame a viable response to the famous Fermi question about ETI—namely, Where are they?  My reply includes a serious timing issue: Conceivably, over the history of our Galaxy, virtually all the bulbs in this extraordinary light fixture eventually illuminate when myriad technologically competent civilizations emerge as evolution aimlessly yet sufficiently twists the many bulbs for each of them to glow—but perhaps only a few light up at any given time.  That is, such a chandelier might never fully shine brilliantly since only a few bulbs simultaneously brighten—maybe only one (or none) illumes at any moment, such as a dim and dirty bulb now signifying us toward the edge of the chandelier. All of which reverie suggests that advanced civilizations might not endure—their technological longevity is brief.

Many SETI enthusiasts lament the fact that no signal has thus far been heard, seen, or otherwise recognized.  SETI proponents are quick to note that today's search strategies resemble hardly more than a romp in a haystack—and that a lack of signal means little as we have so far only briefly sampled the total search domains of space, time, and wavelength.  Among them, I, too, have often mused that only after searching for perhaps a thousand years and still not having detected anything might the observed silence be significant.  Even so, more than five decades of scanning the skies for ETI ought to be sufficient to infer something about the prevalence of smart, long-lived aliens.  This is especially so given the recent rash of exciting discovery of habitable exoplanets, which many colleagues argue increases the probability of making contact, yet by contrast makes lack of contact even more troubling given that we now know the Universe to be rich in favorable extraterrestrial abodes.  For a Universe that seems bio-friendly, absence of evidence for ETI to date implies either that biological intelligence is everywhere slow to evolve as was the case on Earth, or that galactic civilizations don't survive long after achieving technical capability to intercommunicate.  While remaining a staunch SETI supporter, I nonetheless surmise that, at any one time, there are likely very few "needles in the cosmic haystack."

Might we learn something useful from the lack of positive ETI detection?  Is it possible that the absence of an incoming literal signal represents an important figurative signal, which conveys ways and means that our own "bulb in the chandelier" might stay lit longer?

## The Signal

At the Green Bank anniversary meeting, I was assigned to a panel that most participants carefully avoided.  I was asked to critically assess long-standing assumptions regarding SETI in front of esteemed SETI pioneers who had made those same assumptions for decades, and to speculate about what might happen if extraterrestrial intelligence was found—much as I had a decade earlier without much endorsement (Chaisson 2000).  So I trotted out my favorite chandelier analogy and was surprised to hear myself saying that the evident silence could well be telling us that it's time to get our own planetary house in order.

I was even more startled when this inward-focused notion resonated with a couple of administrative assistants who had helped organize the meeting and were eavesdropping on a few of the panels.  These clerical workers as well as a few engineers who had been quietly sitting in on the meeting surrounded me after the session, wanting to know more about how the outward-focused work of the observatory could conceivably aid humankind at a time when global social problems seem to be mounting.  The discussion continued well on into the evening when some cooks and servers joined us after dinner in the laboratory cafeteria. These non-science folks were not interested in contacting ETI as much as learning how the search itself might identify ways that our own civilization on Earth could benefit.

I found this off-line, after-hours conversation quite sobering.  While the distinguished celebrity scientists adjourned for drinks in an upstairs lounge, several observatory support staff had corralled me for a spirited discussion that I never recall having had as a member for nearly two decades of NASA's SETI Science Working Group.  More than ever before, I realized that the lack of ETI signals might actually amount to a "signal" itself—indeed one that might both contain a message and have an impact for us on Earth.  In short, the negative findings thus far for other intelligent life-forms might well be alerting us to get our own worldly act together and thereby enhance society's technological longevity as much as possible—to maximize the "L" factor at the end of the famous Drake equation.

Could the eerie silence from alien worlds actually help to improve *this* world?  Might the null "signal" contain a "message" that our civilization needs to learn to cope with global problems that could harm, reverse, or even extinguish life on Earth?

## The Message

To appreciate any embedded "message" within the null "signal," consider how astrobiology might help us identify practical ways to aid human society at key milestones of our own recent evolutionary trajectory.  Harshly stated though no less true, evolution is all about winners and losers—and regarding life on Earth, the ~99% of all species that are extinct comprise the losers; life's development is not much different, as for example in any typical forested area hardly 20 trees result from 1000 seedlings, most of which die while unable to acquire enough energy.  So might most ETI be losers.  The broader context of cosmic evolution—a synthesis of change writ large among radiation, matter, and life throughout the history of the Universe—can potentially guide us to become a rare winner, hence to increase L and survive longer.  With that in mind, I recently published a study that I never thought I would write—or even ever be able to write—exploring how the cosmology of cosmic evolution might have numerous practical applications to human society, including global warming, smart machines, world economics, and cancer research (Chaisson 2014a).

The grand scenario of cosmic evolution strives to provide a comprehensive worldview of the rise of complex, material systems in the known Universe, from big bang to humankind (Chaisson 2001, 2014b; Dick and Lupisella 2009; Vidal 2014).  Also sometimes termed "universal history," "big history," "epic of evolution," or simply "origins," this interdisciplinary worldview combines complexity science and modern cosmology in order to grant humankind a sense of place in the cosmos.  Cosmic evolution also places the study of life in the Universe into an even larger perspective, providing for astrobiology an effective unifying theme across three preeminent sequential timeframes—Radiation Era → Matter Era → Life Era— the last of these beginning not when life originates rather when it becomes technologically sentient (Chaisson 1987, 2003).

Basically the idea is this: Energy aids the rise of complexity among all material systems, living and non-living.  Energy flow, in particular, through open, organized, non-equilibrium systems seems to be a key facilitator in all aspects of evolution, and may actually drive it, albeit meanderingly, while mixing chance causes with necessary effects. It's a relatively simple idea, dating back to Heraclitus some 25 centuries ago, that is now constantly being tested with new experiments and observations, all in accord with the modern scientific method as scientists of different disciplines go about their business exploring the origin and evolution of myriad diverse systems spanning the Universe. Particularly notable is the rise in complexity of many varied systems over the course of time, indeed dramatically so (with some exceptions) within the past half-billion years since the Cambrian geological period. Remarkably, this energy-centered concept, supported by vast amounts of quantitative data, details how a wealth of complexity emerged, flourished, and grew, despite the entropy-based $2^{nd}$ law of thermodynamics that implies things ought to be breaking down.

Theory, experiment, observation, and computer modeling together affirm that islands of ordered complexity—mainly galaxies, stars, planets, life, and society—are numerically more than balanced by great seas of increasing disorder elsewhere in the larger environments beyond those systems. Emergent complex systems encountered in the cosmic-evolutionary scenario agree quantitatively with the valued principles of thermodynamics, including its celebrated $2^{nd}$ law. Furthermore, an underlying principle, general law, or ongoing process might well aid in the creation, organization, and maintenance of all complex systems everywhere and everywhen, from the early Universe to the present.

Neither absolute energy alone, however, nor merely energy flow are sufficient to explain the origin and evolution of the multitude of systems observed in Nature. Life on Earth is surely more complex than any star or galaxy, yet the latter utilize much more total energy than anything now alive on our planet. Accordingly, I have found that by normalizing energy flows in complex systems by their inherent mass, a more uniform analysis and effective comparison is achieved between and among virtually every kind of system known in Nature. Mass-normalized energy flow, characterized by the term "energy rate density," is a leading candidate underlying a ubiquitous universal process capable of building structures, evolving systems, and creating complexity throughout the cosmos.

Figure 1 plots energy rate density literally "on the same page" for many complex systems, whose numerical values for major systems are compiled in the five bubble inserts. This single graph shows this same quantity consistently and uniformly describing physical, biological, and cultural evolution for an extremely wide range of complex systems spanning ~20 orders of magnitude in spatial dimension and nearly as many in time. For those who prefer words devoid of numbers, a simple "translation" of the figure's rising curves suggests a ranked order of increasingly complex systems across all of time to date:
- mature galaxies are more complex than their dwarf predecessors
- red-giant stars are more complex than their main-sequence counterparts
- eukaryotes are more complex than prokaryotes
- plants are more complex than protists
- animals are more complex than plants
- mammals are more complex than reptiles
- brains are more complex than bodies
- society is more complex than individual humans.

CUP Editor: Please consider reproducing this figure as a whole page. It's the only figure in this paper and contains much information. I could redraw it to match better the page size, once I know the correct aspect ratio. Thanks. --EJC

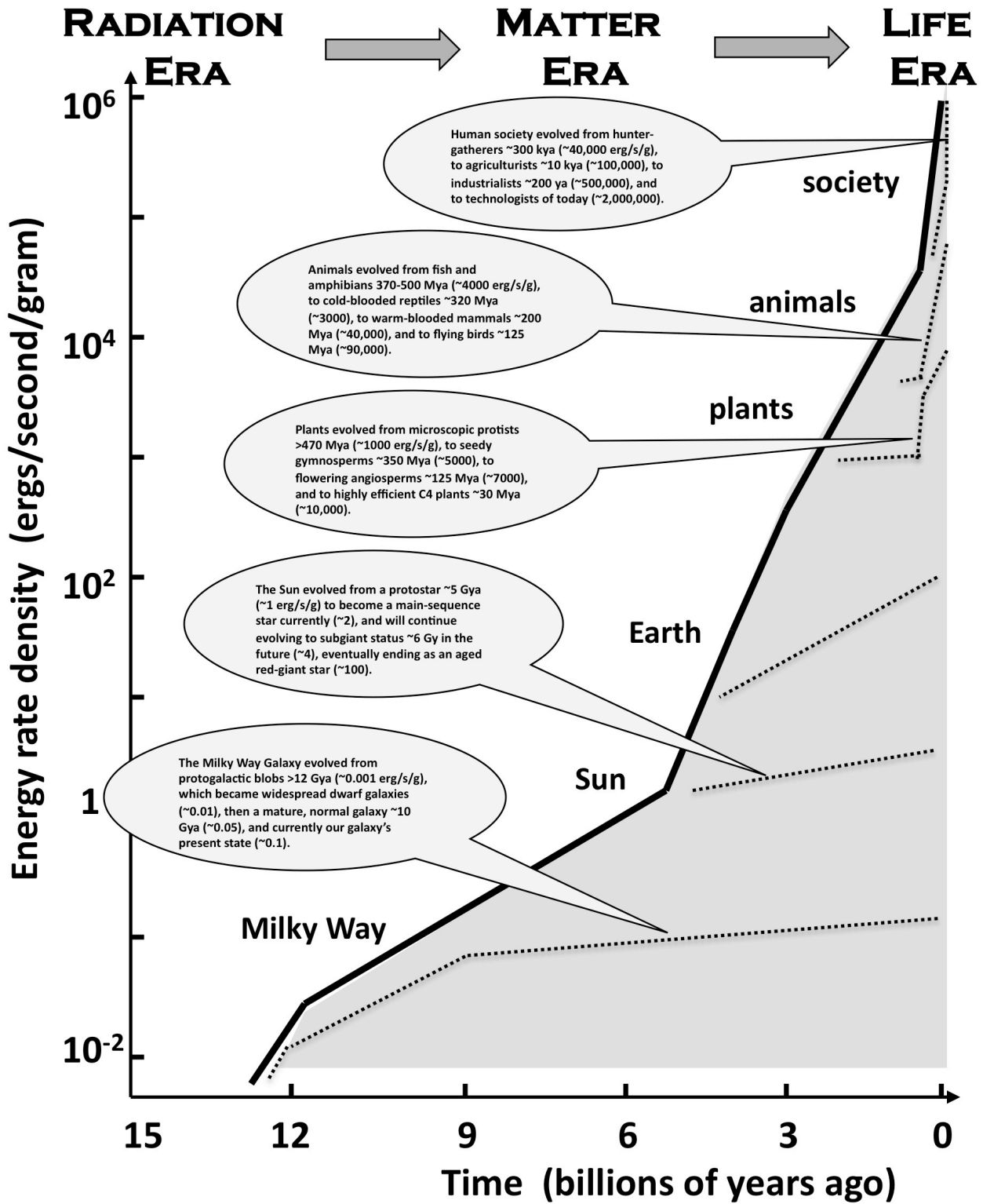

**Figure 1** – Energy rate densities for a wide spectrum of systems observed throughout Nature display a clear trend across ~14 billion years as simple primordial matter changed into increasingly intricate, complex systems. The solid black curve implies an exponential rise in system complexity as cultural evolution (steepest slope at upper right) acts faster than biological evolution (moderate slope in middle part of curve),

which in turn surpasses physical evolution (smallest slope at lower left).  The shaded area includes a huge ensemble of energy rate densities as many varied types of complex systems continued changing and complexifying since their origin; the several dotted black curves delineate notable evolutionary paths traversed by the major systems labeled.  The energy-rate-density values and historical dates plotted here are estimates for specific systems along the evolutionary path from big bang to humankind, namely, our galaxy, star, planet, life, and society, as compiled in the bubble inserts (Chaisson 2014b).  Similar graphs likely pertain to extraterrestrial life-forms, as all complex systems fundamentally hark back to the early Radiation Era, evolve throughout the Matter Era, and potentially enter the Life Era (left to right across top).

What seems inherently attractive is that energy flow as a universal process helps suppress entropy within increasingly ordered, localized systems evolving amidst increasingly disordered, global environments—a process that has arguably governed the emergence and maturity of our galaxy, our star, our planet, and ourselves.  If correct, then energy itself is a principal mechanism of change in the expanding Universe.  And energy rate density is an unambiguous, objective complexity metric enabling us to gauge all organized systems in like manner, as well as to examine how over the course of time some systems evolved to optimally command energy and survive, while others apparently could not and did not.

The graph in Figure 1 encapsulates specifically our own evolutionary trek from big bang to humankind (Chaisson 2001, 2014b; Dick 2012).  It empirically maps the evolutionary path, with neither purpose nor direction, of the Milky Way, Sun and Earth, as well as life and society.  Other graphs of energy rate density would depict the rise of complexity on alien habited worlds, if we only knew of any.  Each graph, for each extraterrestrial civilization, would likely differ, yet minimally as this is a semi-logarithmic scale; its events and phenomena are greatly compressed.  All inanimate complex systems, such as galaxies, stars and planets, populate only the lower parts of such curves; more complex systems, such as animate life-forms and their societies and machines, reside in the upper parts.  The general trend of rising complexity with the advance of cosmic evolution, which is presumed to be a universal process, likely pertains to all galactic civilizations, possibly causing in the above analogy each of their "bulbs" to illuminate—provided their energy-rate-density curves do not turn over or truncate as societal problems impair or destroy their continued development beyond rudimentary technological intelligence.

Close examination of energy rate density in Figure 1 for each type of individual complex system suggests a sharp rise for only limited periods of time, after which some of the curves begin flattening.  Although caution is needed to avoid over-interpreting these data, some (but not all) systems do slow their rate of complexification; they seem to follow a classic, sigmoidal, S-shaped curve—much as microbes do in a petri dish while replicating with fixed food supplies or as human population is expected to plateau later this century.  That is, energy rate densities for a whole array of physical, biological, and cultural systems first increase slowly and then more quickly during their individual evolutionary histories, eventually leveling off throughout the shaded area of Figure 1; if true, then the bold graph in Figure 1 that winds exponentially from lower left to upper right is probably the compound sum of multiple S-curves (Modis 2012).  Viable, complex, surviving systems display no absolute decrease in their energy rate densities, rather merely lessened growth rates and S-shaped inflections as those systems apparently matured.  Ultimately most systems, including unstable stars, stressed species, and inept civilizations, do collapse when they can no longer sustain themselves by optimally managing their energy budgets; such adverse fates are natural, common outcomes of cosmic evolution.

Might part of the "message" contained within the null "signal" be that successful, winning, complex systems, including intelligent life everywhere, are masters of continually increased energy usage—and those that cannot manage to do so are losers? Could this implicit message impact our society on Earth, including our fledgling candidacy for the Life Era?

**The Impact**

The general trend of increased energy utilization in Figure 1 looks and feels like a cosmological imperative. Everlasting evolution and rising complexity may well be hallmarks of Nature, especially in a Universe that expands at an accelerating rate (Chaisson 2001). There is no scientific reason to expect that the main, overall (bold) curve of energy rate density will halt its upward trend or ever turn over. Even if individual systems' (dotted) curves do plateau, such as for maturing galaxies, brainy species or technical gadgets, some other ascendant complex system—such as a symbiosis of meat and machines—will likely vie for ever-higher energy rate density. Not everyone foresees such perpetual growth; some envision an end to high-tech devices lest societies self-destruct (Skrbina 2015), others infer limits to evolutionary complexity and a potential uniqueness for humankind (Conway Morris 2003, 2013).

All things considered and mindful of the extensive data comprising Figure 1, indefinitely rising energy rate densities imply a universal prerequisite for technological civilizations to endure. Societies capable of harnessing optimal amounts of energy consumption per capita—not too little as to starve them and not too much as to destroy them—potentially survive, if only locally and temporarily. If ETI is not as widespread as previously thought—and that seems to be what the lack of SETI success is relating to date—then the rarity of advanced beings implies short longevities probably owing to inability or unwillingness to adapt to larger energy use. A plethora of planetary issues might have caused them to succumb, including global warfare, climate change, asteroid collision, infectious disease, among many other natural or self-inflicted calamities capable of delivering too much or too little energy and thus terminating civilizations. Perhaps complexity does have its limitations after all.

Given that most living species are losers biologically, it is not unreasonable that most galactic civilizations might be losers culturally. They are, quite likely and naturally, cosmically selected out of the population of advanced civilizations, offering a pragmatic solution to the Fermi paradox. The Life Era is perhaps rarely populated by many technological civilizations at any one time, harking back to the few glowing bulbs in the vast chandelier analogy above. Such outcomes are neither special nor predicable. Frequent losers are disfavored while experiencing rapid disaster or slow extinction; they simply make a resource mistake or time their actions incorrectly, both outcomes failing to exploit energy budgets optimally. Rare winners survive favorably perhaps for no other reason than they manage to muddle through each new evolutionary step by effectively utilizing more energy per capita.

Thus, the impact of the "signal" and its "message" for us on Earth conceivably regards increased use of energy to power our civilization going forward. Yet a central issue confronts humanity here and now: How can our technological society get its act together, and quickly too, to foster increased energy use without adversely affecting the environment in which we live? That is, how can humankind's energy rate density continue to rise indefinitely atop Figure 1 without seriously harming our planetary abode? It's a practical matter and a timing issue; like much else in evolution writ large, survival at the threshold of the Life Era is all about energy viability and astute timing.

Society and its machines are currently driven upward and onward by economic growth, which is also an integral part of cultural evolution. The world economy is no different from any other open, complex system with its incoming energy and resources, outgoing products and wastes, and a distinctly non-equilibrium status. Much as for all complex systems, energy is central to the creation, growth, and operation, (as well as demise) of any economy; neither too much nor too little but optimal amounts of energy utilization literally *drive* knowledge creation and product innovation. The bottom line—for this is economics, after all—makes clear that energy is the common currency of economies even more than money or self-interest. Today's most successful businesses are all about speed of production (including design and manufacture) as well as turnaround of new and better products; high-tech communications and intense social networking help to accelerate ideas, research, and development. And nothing speeds things along more than energy, which is at the heart of all complex systems' evolution.

Understandably, social scholars concerned about natural scientists treading on their turf will likely resist notions of non-equilibrium, market gradients, and non-linear dynamics, all of it implying economics (and politics) on the ragged edge of chaos. Yet if we have learned anything from cosmic-evolutionary analyses, realistic economies could be gainfully modeled using principles of complexity science and its underlying thermodynamics (Buchanan 2013; Chaisson 2014a; Gogerty 2014). Whereas orthodox economic modeling (with its supply-demand equilibrium and stable input-output harmony in the marketplace) spawns periodic crises among financial institutions and the nation-states that control them—a classic "heat death" of global markets and likely eventual collapse of technological civilization—non-equilibrium thermodynamics goes about its robust business of guiding energy flows through complex cultural systems, which for the case of human society is ourselves mostly within the vibrant and expansive cities of planet Earth.

Cities are as much a product of cosmic evolution as any star of life-form, and their success and sustainability are closely tied to energy flowing through them. Energy use is already high in the cities where most people live; although cities occupy <1% of Earth's land area, they now house ~55% of humanity and account for ~70% of all global energy usage. Even more energy will likely be needed not only to lift developing nations out of poverty but also to increase the standard of living for the multitudes ( about million people per week) now flocking to cities during the greatest migration in human history. Urban energy metabolism has become a reality on planet Earth, and in today's energy-centered society efficiency gains often ironically dictate yet more energy use, with purported energy savings actually translating into higher consumption—which is why many people who buy cars with good mileage ratings typically drive more and those who are comfortable with smart gadgets tend to own more of them, ultimately often using just as much and sometimes more total energy. As cities grow and complexify, they hunger for more, not less, energy—both in absolute and per-capita amounts; cities are more than the sum of their many residents as economy of scale saves little, if any, energy despite shared infrastructure (Bettencourt *et al.* 2007; Batty 2013; Fragkias *et al.* 2013). Cities exemplify cultural evolution at work in its most rapid and vigorous way to date, yet they are fundamentally no different than other complex systems described by cosmic evolution; humans cluster into cities much like matter assembles into galaxies, stars, and planets, or life itself organizes into bodies, brains, and society. The story is much the same all the way up the rising curve of Figure 1.

Humankind must raise its energy acumen—to think big and adapt broadly to what fundamentally drives human society. That driver is not likely information, the internet, greed, or any other subjective factor that theorists and pundits often preach; rather, all complex systems, including cultured humanity and its smart gadgets, are root-based on energy. Cosmic evolution and its undeniable upward trend near the top of Figure 1 advise copious amounts of additional energy to power society, machines, cities, and the economy as well as their likely symbioses. The implication here is that our descendants' fate is not nearly dependant on more efficiency and less energy as it is on increased energy density. The cosmic-evolutionary narrative asserts that robust energy use, now and forevermore, is likely a cultural requirement if we are to enter the Life Era.

Yet an issue looms large and potentially fatal: Cities are not only voracious users of energy; they are also the largest producers of entropy on the planet. Can humankind increase its energy budget, both in absolute terms and on a per capita basis, without degrading Earth's environment? The $1^{st}$ law of thermodynamics is as inviolable as the $2^{nd}$ law; energy conservation demands that *all* energy utilized from non-renewable sources (not just that inefficiently wasted) eventually dissipates as heat. Alas, heat alone is pollution that can, even absent anthropogenic greenhouse gases, overheat our planet within a few centuries (Chaisson 2008, 2014a; Flanner 2009).

Solar energy is the only natural solution that permits us to circumvent the dilemma that society needs more energy yet cannot tolerate more of its inevitable heat byproduct. Unlike fossil fuels that are dug up on Earth, used on Earth, and warm the Earth internally, the Sun's daily rays land here externally and are already accounted for in our planet's thermal balance. Humankind and its machines can safely utilize more energy without being awash in heat only by adopting solar energy (including its renewable derivatives of wind, water, and waves), plenty of which is available to power economic growth indefinitely; a single hour's dose of incoming radiation from the Sun approximates that now used by civilization in a year. What's more, the Sun's energy seems the only seriously viable way that our technological society can avoid collapse, continue evolving, and ultimately be selected by Nature to endure perhaps indefinitely. This isn't the stance of an astrophysicist looking to the sky for solutions to earthly problems. It's an abbreviation of the single strongest scientific argument to create a solar economy as soon as possible to secure humanity's future well-being.

Such a global solar-based economy would still produce numerous goods and services, but the one vital resource—energy—that underlies all complex systems, including society itself, would no longer be subject to geopolitics, revolutions, or selfishness. Urban energy metabolism can become an earthly virtue, shepherding the structure and function of our cities and their residents without further degrading surrounding environments. Only our parent star can grant us the freedom to employ the needed additional energy required to survive—an idea pondered decades ago (Kardashev 1964) and now surmised as likely much the same everywhere: Technological civilizations anywhere in the Universe that manage to use starlight are naturally selected to endure, and those that don't, aren't. Again, it's all about energy and all about timing. Can civilizations execute both? Have any? Can we?

**The Sum**

More than a half-century of searching for ETI has produced no contact whatsoever. Such a lack of signal, however, might actually constitute a "signal." At the least, many observers suspect that advanced technological civilizations are less abundant than earlier estimates of physicists and astronomers (but probably not most biologists). Astrobiology remains a useful context in which to study life in the Universe, but currently there is no evidence that any intelligent entity has survived long in the Life Era, where resultant system complexity and required energy usage are likely essential, high, and sustained.

Given that ubiquitous change, energy flow, and rising complexity are hallmarks in the emergence of all ordered systems, it might well be that some galactic technological civilizations—perhaps nearly all of them, given the lack of ETI discovery to date—never manage to enhance optimally and expeditiously their energy rate densities. The "message" contained within the "signal" might reveal why ETI is apparently so scarce elsewhere in the Universe: To survive long-term, advanced civilizations need to use the energy of their parent star as quickly as technologically feasible—and those that don't are non-randomly removed from the population of galactic civilizations. In turn, that "signal's message" might also be telling us how to endure on Earth: Embrace the abundant energy of the Sun—not merely use it or lose it, rather use it or lose.

Throughout all of Nature, if complex systems' energy usage is optimum—neither maximized nor minimized, rather within different ranges for different systems of different masses—then those systems have opportunities to survive, prosper, and evolve; if it's not, they are non-randomly eliminated. Civilizations are likely no different. Thus, my considered resolution of the Fermi paradox is mundane and boring: Most, perhaps all, technological societies, much like biological species, are naturally selected out of existence, not because of any dramatic inward failure or outward catastrophe, rather because they simply fail to manage their energy budgets smartly and quickly with the march of time.

Whether civilization on Earth endures or not—the choice is probably ours since technology is society's to wield as we collectively wish—the stars will shine and the galaxies will twirl, with or without sentient beings here or anywhere else. If we do manage to survive into the Life Era, the cosmic-evolutionary narrative can help empower humans and their descendants in countless ways to appreciate not only the importance of utilizing the boundless resource of our parent star, but also how wisely managed and optimally energized complex systems can safeguard the health, wealth, security, and perhaps destiny of humankind.

# References

Batty, M. 2013. *The New Science of Cities*. Cambridge: MIT Press.

Bettencourt, L.M.A., Lobo, L., Helbing, D., Kuhnert, C. and West, G.B. 2007. "Growth, Innovation, Scaling, and the Pace of Life in Cities." *Proc. National Academy of Sciences* 104:7301-7306. http://dx.doi.org/10.1073/pnas.0610172104

Buchanan, M. 2013. *Forecast: What Physics, Meteorology, and the Natural Sciences Can Teach Us About Economics*. New York: Bloomsbury.

Chaisson, E. 1987. *The Life Era: Cosmic Selection and Conscious Evolution*. New York: Atlantic Monthly Press.

Chaisson, E.J. 2000. "Null or Negative Effects of ETI Contact in the Next Millennium." In *When SETI Succeeds*, edited by A. Tough, 59-60, Seattle: Foundation for the Future Proceedings.

Chaisson, E.J. 2001. *Cosmic Evolution: The Rise of Complexity in Nature*. Cambridge: Harvard Univ. Press.

Chaisson, E.J. 2003. "A Unifying Concept for Astrobiology." *International Journal of Astrobiology* 2: 91-101. http://dx.doi.org/10.1017/S1473550403001484

Chaisson, E.J. 2008. "Long-term Global Heating from Energy Usage." *Eos Transactions of the American Geophysical Union* 89:253-254. http://dx.doi.org/10.1029/2008EO280001

Chaisson, E.J. 2014a. "Practical Applications of Cosmology to Human Society." *Natural Science* 6:767-796. http://dx.doi.org/10.4236/ns.2014.610077

Chaisson, E.J. 2014b. "The Natural Science Underlying Big History." *The Scientific World Journal* vol. 2014, Article ID 384912, 41 pages. http://dx.doi.org/10.1155/2014/384912

Conway Morris, S. 2003. *Life's Solution: Inevitable Humans in Lonely Universe*. Cambridge: Cambridge Univ. Press.

Conway Morris, S. 2013. "Life: the final frontier for complexity?" In *Complexity and the Arrow of Time*, edited by C. Lineweaver, P. Davies and M. Ruse, 135-161, Cambridge: Cambridge Univ. Press.

Dick, S.J. 2012. "Critical Issues in History, Philosophy and Sociology of Astrobiology." *Astrobiology* 12:906-927.

Dick, S.J. and Lupisella, M.L. (eds.). 2009. *Cosmos & Culture: Cultural Evolution in a Cosmic Context*. Washington: NASA SP-2009 4802.

Flanner, M.G. 2009. "Integrated anthropogenic heat flux with global climate models." *Geophysical Research Letters* 36:L02801-L02805. http://dx.doi.org/10.1029/2008GL036465

Fragkias, M., Lobo, J., Strumsky, D. and Seto, K.C. 2013. "Does Size Matter? Scaling of $CO_2$ Emissions and US Urban Areas." *PLoS ONE* 8:1-6. http://dx.doi.org/10.1371/journal.pone.0064727

Gogerty, N. 2014. *The Nature of Value*. New York: Columbia Univ. Press.

Kardashev, N. 1964. "Transmission of Information by Extraterrestrial Civilizations." *Soviet Astronomy* 8:217–221.

Modis, T. 2012. "Why the Singularity Cannot Happen." In *Singularity Hypotheses*, edited by A.H. Eden *et al.*, 311-340, Berlin: Springer-Verlag.

Skrbina, D. 2015. *The Metaphysics of Technology*. New York: Routledge.

Vidal, C. 2014. *The Beginning and the End*, Switzerland: Springer International Publishing.